\documentclass[fleqn,twoside]{article}
\usepackage{espcrc2}

\usepackage{graphicx}
\usepackage[figuresright]{rotating}

\newcommand{\AmS}{{\protect\the\textfont2
  A\kern-.1667em\lower.5ex\hbox{M}\kern-.125emS}}

\newcommand{\bsg}{\ensuremath{B\to X_s \gamma}} 
\newcommand{\fb}{fb\ensuremath{^{-1}}}
\newcommand{\BB}{\ensuremath{B{\overline B}}}
\newcommand{\sll}{\ensuremath{b\to s\ell^+\ell^-}}
\newcommand{\Kll}{\ensuremath{B\to K^{(*)}\ell^+\ell^-}}
\newcommand{\mes}{\ensuremath{m_{ES}}}
\newcommand{\ellell}{\ensuremath{\ell^+\ell^-}}
\newcommand{\ee}{\ensuremath{e^+e^-}}
\newcommand{\mm}{\ensuremath{\mu^+\mu^-}}
\newcommand{\Kmaybestar}{\ensuremath{K^{(*)}}}
\newcommand{\ksll}{\ensuremath{K^*\ellell}}
\newcommand{\fl}{\ensuremath{F_L}}
\newcommand{\afb}{\ensuremath{A_{FB}}}
\newcommand{\bkll}{\ensuremath{B\to K\ellell}}
\newcommand{\bpll}{\ensuremath{B\to \pi\ellell}}
\newcommand{\bknn}{\ensuremath{B\to K^{(*)}\nu\overline{\nu}}}
\newcommand{\btn}{\ensuremath{B\to\tau\nu}}

\usepackage{relsize}
\def\babar{\mbox{\slshape B\kern-0.1em{\smaller A}\kern-0.1em
    B\kern-0.1em{\smaller A\kern-0.2em R}}}

\title{Radiative and Leptonic $B$-meson Decays from the B-factories}

\author{John Walsh\address{Istituto Nazionale di Fisica Nucleare \\
         Largo Pontecorvo, 3 \\
         Pisa, Italy}, on behalf of the \babar\ and Belle Collaborations}

\begin{document}

\begin{abstract}
Radiative and leptonic decays of $B$-mesons represent an excellent
laboratory for the search for New Physics.  I present here recent
results on radiative and leptonic decays from the Belle and
\babar\ collaborations.

\end{abstract}

\maketitle

\section{Radiative Penguin and Leptonic $B$-meson Decays}

Radiative penguin decays of $B$-mesons, in which a $b$ quark transitions to an $s$
or $d$ quark accompanied by either a photon or a pair of charged
leptons (Fig. 1), are a sensitive probe of New Physics (NP) beyond the
Standard Model (SM).  These flavour changing neutral current decays
are forbidden at tree level in the SM and hence strongly suppressed.
In many NP scenarios these decays appear at the one-loop level,
{\em i.e.}, at the same order as the SM processes. Hence, the contributions
of NP to branching ratios and asymmetries can be as large as the SM
contributions, making these decays a good hunting ground for New
Physics. 
%
\begin{figure}[bh]
\begin{center}
\vspace*{-2em}
\includegraphics[width=.6\linewidth]{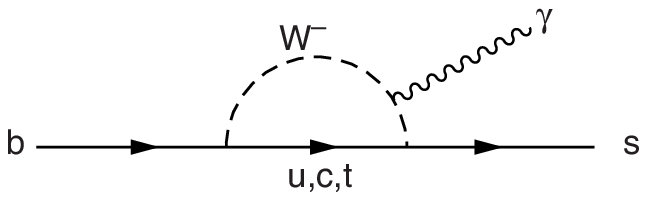} \\
\includegraphics[width=.6\linewidth]{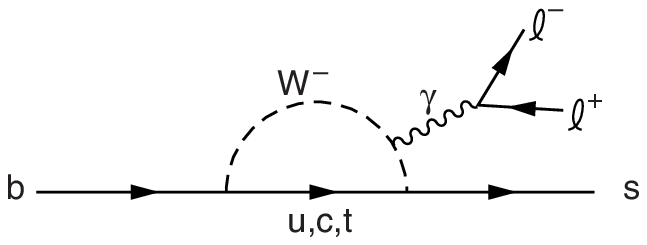} \\
\end{center}
\vspace*{-3em}
  \caption{\label{diagrams}
    Feynman diagrams for radiative penguin processes. Additional diagrams with
    $Z^0$ and $W^\pm$ propagators not shown. 
}
\end{figure}
%

Leptonic $B$-meson decays, in which all final state particles are either
charged leptons or neutrinos, are also often sensitive to
the presence of New Physics.  As with the radiative penguin decays,
virtual heavy particles predicted by NP can contribute to leptonic
decays.

In this report, I discuss recent measurements of radiative penguin and
leptonic $B$-meson decays performed at the \babar~\cite{BabarDetector} and Belle~\cite{BelleDetector} 
experiments.

\section{Inclusive $B \to X_s \gamma$}

The branching fraction of the Standard Model inclusive \bsg\ process
is calculable to a precision of around 8\%~\cite{SM-bsg}:
$$ {\cal B}(B\to X_s \gamma)|_{E_\gamma>1.6
  \rm{GeV}}=(3.15\pm0.23)\times 10^{-4}, 
$$
where the rate is computed above the conventional lower limit of 1.6
GeV.  The good precision of this calculation makes this process a
powerful one in the search for NP and considerable experimental
efforts have been put into making precise measurements of the inclusive
rate.

Inclusive measurements are generally experimentally more difficult
than reconstructing exclusive states: the lack of kinematic constraints
leads to large backgrounds that require careful treatment.  Recent
measurements of \bsg\ have relied on reducing the background from
light quark events by tagging the other $B$-meson in the event,
either by reconstructing it fully~\cite{bsg-breco} or by identifying
the lepton coming from a semi-leptonic decay~\cite{bsg-sl}.  

Ideally, one would like to measure the branching fraction over the
full photon energy range, but experimental considerations require a
minimum photon energy, generally in the neighborhood of 1.8-1.9 GeV.
Since the extrapolation down to 1.6 GeV introduces some
model-dependence, experimentalists strive to keep this photon energy
requirement as low as possible.

Belle has recently reported a new preliminary measurement of \bsg,
based on 605 \fb\ of data, where they do not require any sort of
tagging of the other $B$-meson~\cite{bsg-belle}.  The advantage
of this approach is the large gain in efficiency. The large
backgrounds from continuum (light quark) events are suppressed using
two Fisher discriminants, which exploit the topological differences in
signal and background decays. Additional background suppression is
achieved by directly reconstructing and vetoing $\pi^0 \to \gamma
\gamma$ and $\eta \to \gamma \gamma$ decays.

The remaining continuum background is estimated by analyzing data
taken off-resonance, where no $B$-meson decays are present.
Backgrounds from \BB\ events are estimated using simulated
events. Studies of control samples of real data allow for the correction
of the Monte Carlo estimates for the major contributions to the
\BB\ backgrounds.

\begin{figure}
\begin{center}
\includegraphics[width=.45\textwidth]{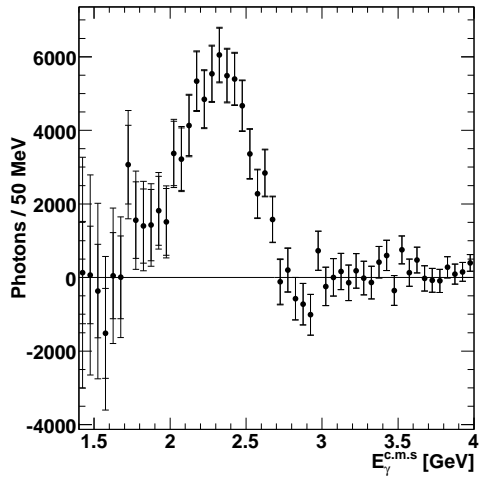} \\
\end{center}
 \vspace*{-3em}
  \caption{\label{bsg}
    The extracted photon energy spectrum for Belle's measurement $B\to X_{s,d}\gamma$. 
    The two error bars show the statistical and total errors.
}
\end{figure}

The resulting background-subtracted photon spectrum for \bsg\ decays
is shown in Fig.~\ref{bsg}. 
%
%
The partial branching fraction, measured in the range $1.7 <
E_\gamma < 2.8$ GeV, where $E_\gamma$ is measured in the B-meson rest
frame, is found to be
$$
  \mathcal{B}\left( B\to X_s \gamma \right) = \left( 3.31 \pm 0.19
  \pm 0.37 \pm 0.01 \right )\times10^{-4}
$$ 
where the errors are statistical, systematic and the uncertainty
arising from translating from the center-of-mass frame to the
$B$-meson rest frame.  The value of the lower energy cut, 1.7 GeV, is
the lowest achieved thus far in a measurement of inclusive \bsg.
These results are preliminary. 

\section{Analysis of $B\to K^{(*)}\ell^+\ell^-$}

The decays $B\to K^{(*)}\ell^+\ell^-$ are exclusives modes of the $b\to
s\ell^+\ell^-$ process, which is analogous to the $b\to s\gamma$
decay, with the photon producing a lepton pair.  In the $b\to s\ellell$ case,
however, there are electroweak contributions, with the $Z^0$ boson
replacing the photon, plus an additional ``box'' diagram involving
$W^\pm$ bosons.  The contributing amplitudes are expressed in terms of
hadronic form factors and effective Wilson coefficients
$C_7^{\rm{eff}}$, $C_9^{\rm{eff}}$ and $C_{10}^{\rm{eff}}$,
corresponding to the electromagnetic penguin diagram, and the vector
and axial-vector part of the $Z^0$ and $W^+W^-$ box diagrams,
respectively~\cite{sll-theory}.

The rate for \sll\ is quite small and experimental efforts have mostly
been expended on studying exclusive states, where the background can
be kept to a low level.  This means that branching fraction
measurements become less important, since the uncertainties on the SM
predictions for exclusive \Kll\ are on the order of 30\%~\cite{Kll-theory}.

The presence of the lepton pair in the final state gives rise to new
observables, compared to \bsg, such as the forward-backward lepton
asymmetry and the $K^*$ polarization fraction.  Additional quantities
that are sensitive to NP, such as the direct CP and isospin
asymmetries and the lepton flavour ratio, tend to have more precise SM
predictions than the branching fractions, because some uncertainties
cancel when taking ratios. In general, the effect of NP on these
quantities will be a function of $q^2 \equiv m_{\ellell}^2$, and
measurements are performed in bins of $q^2$ where possible.

Recent measurements of these quantities have been performed at
\babar~\cite{kll-rates,kll-afb}. The following preliminary results are based
on a data sample of 384 million $B\overline{B}$ pairs. 

The decays \Kll\ are reconstructed in ten different final states
containing an $e^+e^-$ or $\mu^+\mu^-$ pair, together with a $K_s^0$,
$K^+$ or $K^*(892)$ candidate, where $K^*(892)$ is reconstructed in
the $K^+\pi^-$, $K^+\pi^0$ and $K_s^0\pi^+$ modes.  The primary
backgrounds from semileptonic decays of $B$ and $D$ mesons are
suppressed using multivariate techniques, while events
containing $J/\psi$ or $\psi^\prime$ decays are vetoed explicitly. 

The events are divided into a low $q^2$ region ($0.1 < q^2 < 7.02$ GeV$^2$/c$^4$) 
and a high $q^2$ region ($q^2 > 10.24$ GeV$^2$/c$^4$).

$B$-meson candidates are identified using the kinematic variables
$\mes=\sqrt{s/4 -p^{*2}_B}$ and $\Delta E = E_B^* - \sqrt{s}/2$, where
$p^*_B$ and $E_B^*$ are the $B$ momentum and energy in the
$\Upsilon(4S)$ center-of-mass (CM) frame, and $\sqrt{s}$ is the total
CM energy.  Typically, a selection is made on $\Delta E$ (either $\pm
40$ MeV or $\pm 50$ MeV, depending on the $q^2$ region) and a fit to
\mes\ is used to extract the signal yield.

\begin{figure}[tb]
\begin{center}
\includegraphics[width=0.45\textwidth]{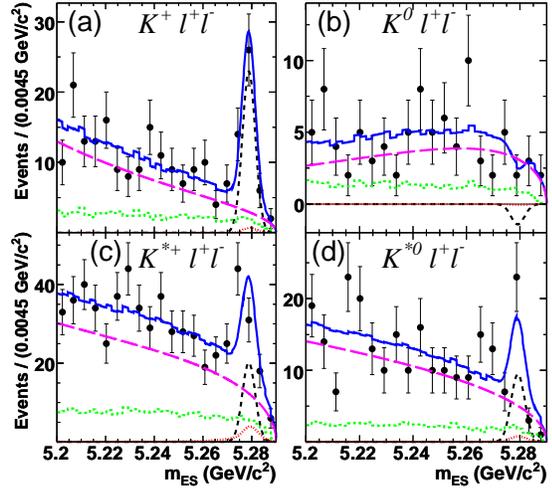}
\vspace*{-2em}
\caption{Distributions of \mes\ with fit results 
in the low $q^2$ region for the \babar\ 
\Kll\ analysis. 
Total fit [solid],
combinatoric background [long dash],
signal [medium dash],
hadronic background [short dash],
peaking background [dots].}
\label{fig:alliso}
\end{center}
\end{figure}

Example \mes\ fits are depicted in Fig.~\ref{fig:alliso}, which shows the
distributions for neutral and charged $K$ and $K^*$ channels in
the low-$q^2$ region. 

\subsection{Direct CP, Isospin and Lepton Flavour Asymmetries}

The extraction of event yields described above allows the measurement
of several asymmetries that are sensitive to the presence of New
Physics.  

The direct CP asymmetry
\begin{displaymath}
A_{CP} \equiv
\frac
{{\cal B}(\overline{B} \rightarrow \overline{K}^{(*)}\ellell) - {\cal B}(B \rightarrow K^{(*)}\ellell)}
{{\cal B}(\overline{B} \rightarrow \overline{K}^{(*)}\ellell) + {\cal B}(B \rightarrow K^{(*)}\ellell)}
\end{displaymath}
is expected to be $O(10^{-3})$ in the SM, but new physics at the electroweak
scale could produce a significant enhancement~\cite{Bobeth:2008hp}.
The results found for this analysis are reported in
Table~\ref{tab:acpressys}.  We find no evidence for a deviation from
the Standard Model. 
%
%
\begin{table}[b!]
\centering
\caption{$A_{CP}$ results from \babar. The high $q^2$ region, not shown here 
for lack of space, appears in~\cite{kll-rates}. The quoted
errors are statistical and systematic, respectively.}
\label{tab:acpressys}
{\footnotesize
\begin{tabular}{lcc}
\hline \hline
Mode & combined $q^2$ & low $q^2$ 
\\ \hline
   $K^+\ellell$ & $-0.18_{-0.18}^{+0.18}\pm0.01$ & $-0.18_{-0.19}^{+0.19}\pm0.01$  \\
$K^{*0}\ellell$ & $0.02_{-0.20}^{+0.20}\pm0.02$  & $-0.23_{-0.38}^{+0.38}\pm0.02$  \\
$K^{*+}\ellell$ & $0.01_{-0.24}^{+0.26}\pm0.02$  & $0.10_{-0.24}^{+0.25}\pm0.02$   \\
   $K^*\ellell$ & $0.01_{-0.15}^{+0.16}\pm0.01$  & $0.01_{-0.20}^{+0.21}\pm0.01$   \\
\hline \hline
\end{tabular}
}
\end{table}
%

In the Standard Model the ratio of rates to di-muon and di-electron final states
\begin{eqnarray}
R_{\Kmaybestar} \equiv
\frac
{{\cal B}(\Kmaybestar \mu^+\mu^-)}
{{\cal B}(\Kmaybestar e^+e^-)}
\end{eqnarray}
is very close to one~\cite{Hiller:2003js}. In some NP models, such as
supersymmetry,  this ratio can be altered significantly. Deviations
from unity as large as $\sim 10\%$ are possible for large values of
$\tan \beta$~\cite{Yan:2000dc}.

In the region $q^2 < (2m_{\mu})^2$, where only the $e^+e^-$ modes are
allowed, there is a large enhancement of $K^* e^+e^-$ due to a
$1/q^2$ scaling of the photon penguin.  The expected SM value of
$R_{K^*}$ including this region is 0.75~\cite{Hiller:2003js}. The
results for $R_{\Kmaybestar}$, shown in Table~\ref{tab:emuressys},
indicate no deviation from the Standard Model.
%
%
\begin{table}
\caption{$R_{K^{(*)}}$ results in each $q^2$ region.
The extended (``ext.'') regions are relevant only for $R_{K^*}$.
The errors are statistical and systematic, respectively.}
\centering
{\footnotesize
\begin{tabular}{lcc}
\hline \hline
$q^2$ Region  & $R_{K^*}$                      & $R_K$
\\ \hline
combined          & $1.37_{-0.40}^{+0.53}\pm 0.09$ & $0.96_{-0.34}^{+0.44}\pm 0.05$ \\
ext. comb.     & $1.10_{-0.32}^{+0.42}\pm 0.07$ & ---                            \\
low               & $1.01_{-0.44}^{+0.58}\pm 0.08$ & $0.40_{-0.23}^{+0.30}\pm 0.02$ \\
ext. low          & $0.56_{-0.23}^{+0.29}\pm 0.04$ & ---                            \\
high              & $2.15_{-0.78}^{+1.42}\pm 0.15$ & $1.06_{-0.51}^{+0.81}\pm 0.06$ \\
\hline \hline
\end{tabular}
}
\label{tab:emuressys}
\end{table}
%

The CP-averaged isospin asymmetry:
\begin{eqnarray*}
\lefteqn{A^{K^{(*)}}_{I} \equiv} \\ 
 & & \frac
{{\cal B}(B^0\to K^{(*)0}\ellell) - r {\cal B}(B^{\pm} \to K^{(*)\pm}\ellell)}
{{\cal B}(B^0 \to K^{(*)0}\ellell) + r {\cal B}(B^{\pm} \to K^{(*)\pm}\ellell)},
\end{eqnarray*}
\noindent
where $r = \tau_0/\tau_+=1/(1.07\pm 0.01)$ is the ratio of
the $B^0$ and $B^+$ lifetimes~\cite{hfag},
has a SM expectation of $6-13\%$ as $q^2 \rightarrow 0$ GeV$^2$~\cite{Feldmann:2002iw}.
A calculation of the predicted $K^{*+}$ and $K^{*0}$
rates integrated over the low $q^2$
region gives $A^{K^{*}}_{I} = -0.005 \pm 0.020$~\cite{beneke05,Feldmann:priv2008}.

The results for the isospin asymmetry are presented in
Table~\ref{tab:isoressys}.  While we find no significant asymmetry in
the high or combined $q^2$ regions, we do measure a large asymmetry in
the low $q^2$ region.  The origin of this non-zero asymmetry can be
observed in Fig.~\ref{fig:alliso} above, where fewer events are found in
the neutral modes compared to the charged modes.  Combining all modes
in the low $q^2$ region, we find $A_{I}^{\Kmaybestar} =
-0.64^{+0.15}_{-0.14} \pm 0.03$. Including systematics, this is a $3.9
\sigma$ difference from a null $A_{I}^{\Kmaybestar}$ hypothesis.
%
\begin{table}
\centering
\caption{$A_{I}^{\Kmaybestar}$ results from \babar.
The high $q^2$ region, not shown here 
for lack of space, appears in~\cite{kll-rates}. The errors are statistical and systematic, respectively.
The last table row shows $K^*\ee$ results for the extended regions.}
{\footnotesize
\begin{tabular}{lcc}
\hline \hline
Mode          & combined $q^2$  & low $q^2$ 
\\ \hline
$K\mm$ & $0.13_{-0.37}^{+0.29}\pm0.04$         &  $-0.91_{\mathrm{-\infty}}^{+1.2}\pm0.18$   \\
$K\ee$ & $-0.73_{-0.50}^{+0.39}\pm0.04$        & $-1.41_{-0.69}^{+0.49} \pm 0.04$            \\
$K\ellell $ & $-0.37_{-0.34}^{+0.27}\pm0.04$   & $-1.43_{-0.85}^{+0.56} \pm 0.05$            \\
$K^*\mm$  & $-0.00_{-0.26}^{+0.36}\pm0.05$     & $-0.26_{-0.34}^{+0.50} \pm0.05$             \\
$K^*\ee$  & $-0.20_{-0.20}^{+0.22}\pm0.03$     & $-0.66_{-0.17}^{+0.19} \pm0.02$             \\
$K^*\ellell$  & $-0.12_{-0.16}^{+0.18}\pm0.04$ & $-0.56_{-0.15}^{+0.17} \pm0.03$             \\ \hline
$K^*\ee$  & $-0.27_{-0.18}^{+0.21}\pm 0.03$    & $-0.25_{-0.18}^{+0.20} \pm 0.03$           \\
\hline \hline
\end{tabular}
}
\label{tab:isoressys}
\end{table}

This large isospin asymmetry is unexpected in the Standard Model, which predicts
essentially zero asymmetry integrated over the low $q^2$ region. 

These results are preliminary.

\subsection{Angular analysis of $K^*\ellell$ decays}

It is also possible to probe the effects of New Physics by analyzing the 
angular distributions of the decay products in \ksll\ decays as a function of $q^2$~\cite{afb-theory}. 
In particular, the $K^*$ polarization fraction can be determined from the
distribution of the angle $\theta_K$
\begin{displaymath}
\frac{3}{2} \fl \cos^2\theta_K + \frac{3}{4}(1-\fl)(1-\cos^2\theta_K)
\end{displaymath}
where $\theta_K$ is the angle between the $K$ and $B$ directions in the $K^*$ rest frame.

Furthermore, the lepton forward-backward asymmetry, \afb, can be determined from the distribution of
the angle $\theta_\ell$
\begin{eqnarray*}
\lefteqn{{{3}\over{4}}\fl (1-\cos^2\theta_\ell)} \\ 
& & + {{3}\over{8}}(1-\fl )(1+\cos^2\theta_\ell) + \afb \cos\theta_\ell.
\end{eqnarray*}
Here $\theta_\ell$ is the angle between the $\ell^+$ ($\ell^-$) and the $B$ ($\overline{B}$) 
direction in the \ellell\ rest frame. 

\begin{figure}[tb!]
\begin{center}
\includegraphics[width=1.0\linewidth]{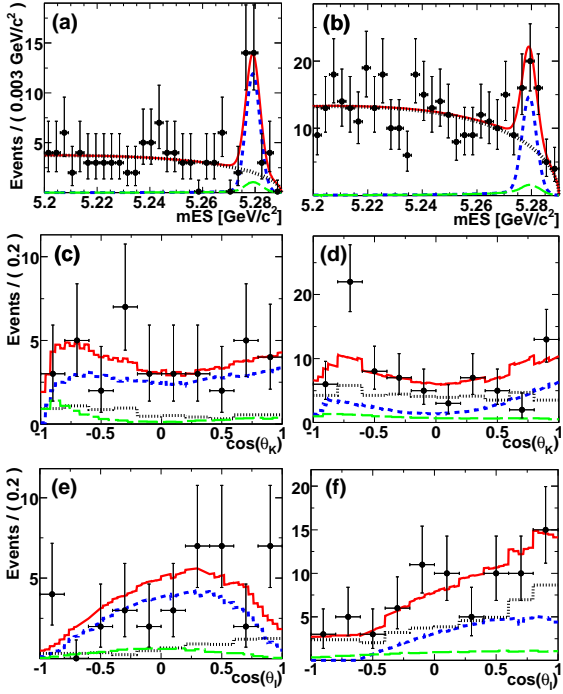}
\vspace*{-3em}
\caption{$K^*\ellell$ fits from \babar:
(a) low $q^2$ \mes,
(b) high $q^2$ \mes,
(c) low $q^2$ $\cos\theta_K$,
(d) high $q^2$ $\cos\theta_K$,
(e) low $q^2$ $\cos\theta_\ell$,
(f) high $q^2$ $\cos\theta_\ell$;
with combinatorial (dots) and
peaking (long dash) background,
signal (short dash) and total (solid)
fit distributions superimposed
on the data points.}
\label{fig:fitplots}
\end{center}
\end{figure}

\begin{figure}[tb!]
\begin{center}
\includegraphics[width=1.0\linewidth]{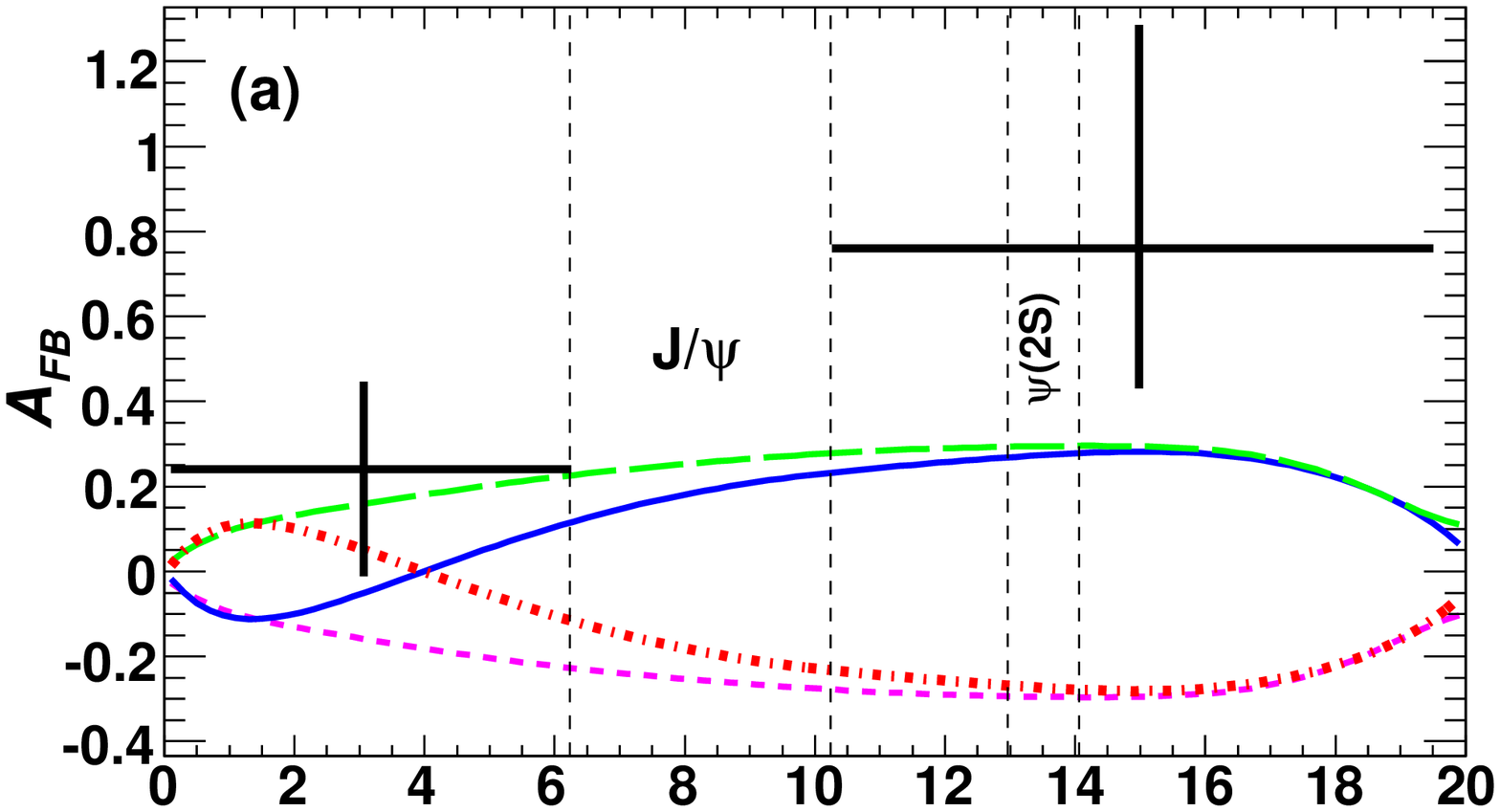}
\includegraphics[width=1.0\linewidth]{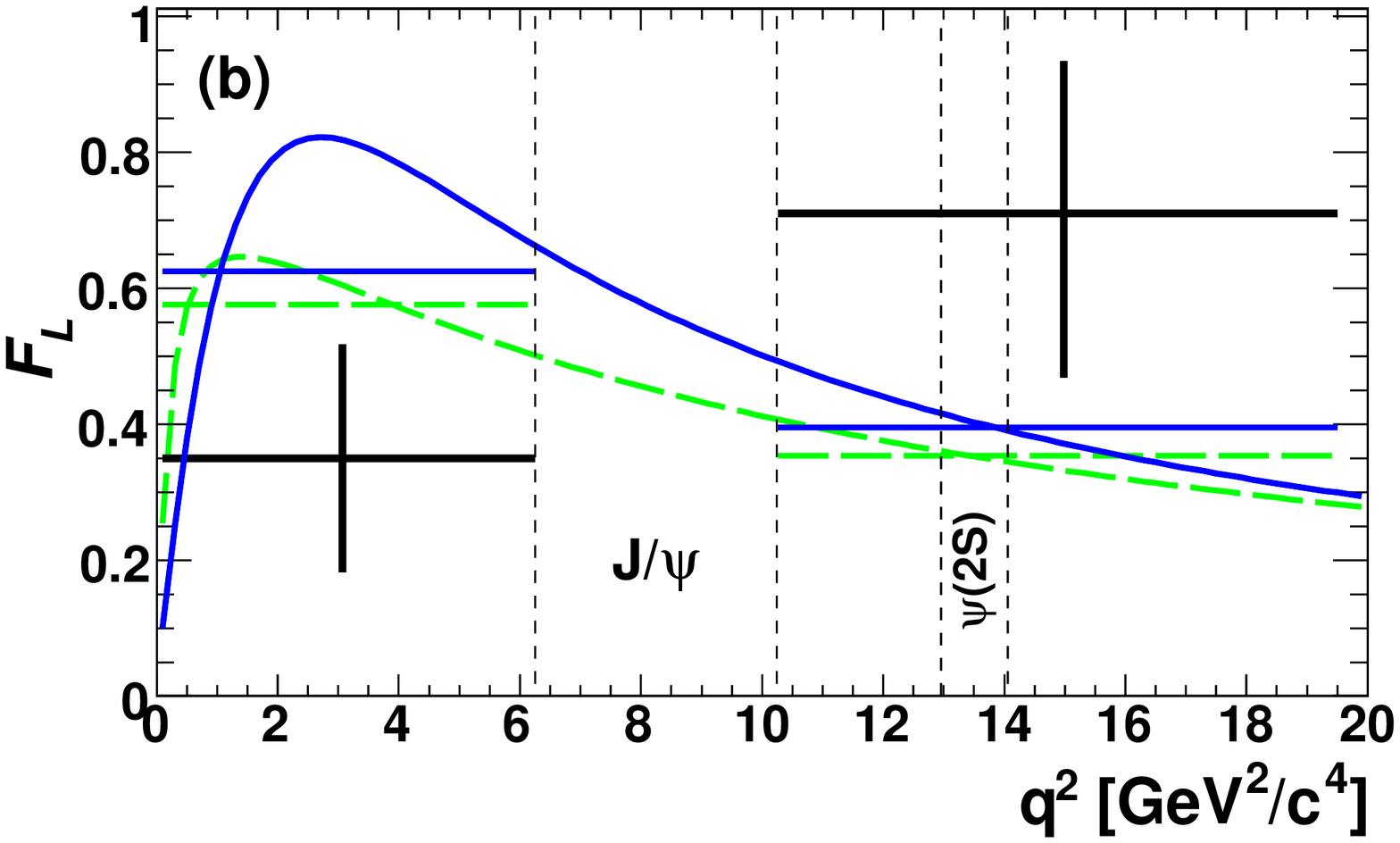}
\vspace*{-3em}
\caption{Plots of \babar's results for (a) $\afb$ and (b) $F_{L}$
for the decay $B\to K^* \ell^+\ell^-$ showing comparisons
with SM (solid); $C_7 = -C_7^{SM}$ (long dash);
$C_9 C_{10} = -C_9^{SM} C_{10}^{SM}$ (short dash);
$C_7 = -C_7^{SM}, C_9 C_{10} = -C_9^{SM} C_{10}^{SM}$ (dash-dot).
Statistical and systematic errors are added in quadrature.
Expected $F_{L}$ values integrated over each $q^2$ region are also shown.
The $F_{L}$ curves with $C_9 C_{10} = -C_9^{SM} C_{10}^{SM}$
are nearly identical to the two curves shown.}
\label{fig:newafbfl}
\end{center}
\end{figure}

Fits to the $K^*\ellell$ sample are shown in Fig.~\ref{fig:fitplots}
and the extracted values of \fl\ and \afb\ are shown in
Table~\ref{tab:kstll} and plotted in Fig.~\ref{fig:newafbfl}.  Results
are consistent with the Standard Model, and the \afb\ results
exclude a wrong-sign $C_9C_{10}$ from purely right-handed weak
currents at 3$\sigma$ significance.

These results are preliminary.
\begin{table}[tb!]
\centering
\caption{Results for the fits to the $K\ellell$ and $K^*\ellell$ samples.
$N_S$ is the number of signal events in the $\mes$ fit.
The quoted errors are statistical only.}
{\footnotesize
\begin{tabular}{lcccc} \hline\hline
Decay & $q^2$  & $N_S$ &  $\fl$ & $\afb$ \\ \hline \vspace{-.1in}\\ \vspace{.04in}
$K\ellell$ & low & $26.0\pm 5.7$ & & $+0.04^{+0.16}_{-0.24}$\\ \vspace{.04in}
& high & $26.5\pm 6.7$ &  & $+0.20^{+0.14}_{-0.22}$ \\ \hline\vspace{-.1in}\\ \vspace{.04in}
$K^*\ellell$ &  low  & 27.2 $\pm$ 6.3 &  $0.35\pm 0.16$ & $+0.24^{+0.18}_{-0.23}$ \\\vspace{.04in}
& high  & 36.6 $\pm$ 9.6 & $0.71^{+0.20}_{-0.22}$ & $+0.76^{+0.52}_{-0.32}$ \\
\hline\hline
\end{tabular}
}
\label{tab:kstll}
\end{table}

\section{Search for $B\to\pi\ellell$}

The rare mode \bpll\ is the $b\to d\gamma$ analogue to the
\bkll\ decay.  The Standard Model rate for \bpll\ is thus suppressed
relative to the \bkll\ mode by the factor $|V_{td}/V_{ts}|^2\sim
0.04$; the SM predictions for the branching fractions are $3.3\times
10^{-8}$ and $1.7\times 10^{-8}$ for the charged and neutral modes,
respectively.

Belle has recently performed a search for \bpll\ on a data sample of 605 \fb,
reconstructing both the neutral and charged modes, with $\ell$ being
$\mu$ or $e$~\cite{pill-belle}.  A Fisher discriminant using 16 event shape
variables suppresses background from the continuum, while events
containing $J/\psi$ or $\psi^\prime$ decays are vetoed explicitly. A
fit to the beam-constrained mass is used to extract the yields. 

No significant signal was found. 
These preliminary results, shown in Table~\ref{tab:pill}, indicate that current
sensitivity is approaching the Standard Model prediction for the
charged mode, although there is still a ways to go for the neutral
mode. 
\begin{table}[t!]
\caption{A summary of the signal yields ($N_s$), reconstruction efficiencies ($\epsilon$), 
and 90\% confidence level upper limits (U.L.) for $B\to\pi\ellell$.
}
\label{tab:pill}
\begin{center}
\begin{tabular}{lccc}
\hline
Mode~               & $N_s$ & $\epsilon$ (\%)      & U.L. ($10^{-8}$) \\ 
\hline
$B^+ \to \pi^+\mu^+\mu^-$   & $0.5^{+2.8}_{-1.9}$~ & 13.1 & 6.9~ \\
$B^+ \to \pi^+e^+e^-$       & $1.4^{+3.2}_{-2.3}$~ & 13.8 & 8.0~ \\
$B^+ \to \pi^+\ell^+\ell^-$ & -                  ~ & - & 4.9~ \\
\hline
$B^0 \to \pi^0\mu^+\mu^-$   & $5.1^{+4.2}_{-3.3}$~ & 9.6 & 18.4~ \\
$B^0 \to \pi^0e^+e^-$       & $2.7^{+5.2}_{-4.0}$~ & 7.4 & 22.7~ \\
$B^0 \to \pi^0\ell^+\ell^-$ & -                    & - & 15.4~ \\
\hline
$B   \to \pi\ell^+\ell^-$   & -                    & - & 6.2~ \\
\hline
\end{tabular}
\end{center}
\end{table} 
 
\section{Searches for $B\to K^{(*)}\nu\overline{\nu}$}

The mode \bknn\ is similar to the $B\to K^{(*)}\ellell$ mode, but with the 
charged lepton pair replaced by a pair of neutrinos.  Since the neutrinos are 
not detected experimentally, $B\to K^*\ellell$ observables such as \afb\ are not 
measurable and the focus on this channel has been to measure its branching fraction. 

The SM branching fractions are predicted to be
$(3.8^{+1.2}_{-0.6})\times 10^{-6}$ and $(1.3^{+0.4}_{-0.3})\times
10^{-5}$ for the $K\nu\overline{\nu}$ and $K^*\nu\overline{\nu}$
channels, respectively~\cite{knn-theory}. These are larger than for
the corresponding $B\to K^{(*)}\ellell$ modes, but the missing
neutrinos present considerable experimental difficulties, leading to
lower sensitivities for this channel.  The neutrinos' momentum is
typically inferred by reconstructing all the other particles in the
event.  When the other $B$-meson is fully reconstructed, the
signal decay yields a single charged track (or $K^*$ candidate) and
significant missing energy.

\babar\ has recently performed searches for $B\to K\nu\overline{\nu}$
and $B\to K^*\nu\overline{\nu}$ reconstructing the other $B$-meson in
the event in a semileptonic decay mode, $B\to D\ell\nu(X)$.
Multivariate discriminators are used to reduce the background. For the
$B\to K^*\nu\overline{\nu}$ mode, a variable $E_{\rm{extra}}$ is
constructed: it is the sum of the leftover energy in the event,
excluding the particles used to reconstruct the tag-side $B$-meson and
the signal $K^*$ candidate.  An excess of events at $E_{\rm{extra}}
\sim 0$ would be an indication of $B\to K^*\nu\overline{\nu}$ decays.
For the $B\to K\nu\overline{\nu}$ channel, the multivariate
discriminator is used to isolate the signal.  For both channels, there
is no evidence of a significant signal and the derived 90\% confidence
level upper limits are found to be
\begin{eqnarray*}
\mathcal{B}(B^+\to K^{*+}\nu\overline{\nu}) & < & 9   \times 10^{-5} \\
\mathcal{B}(B^0\to K^{*0}\nu\overline{\nu}) & < & 21  \times 10^{-5} \\
\mathcal{B}(B^+\to K^+\nu\overline{\nu})    & < & 4.2 \times 10^{-5}
\end{eqnarray*}
The limits for $B\to K^*\nu\overline{\nu}$ are better than previously
report results~\cite{belle-knn}, although we are still fairly far from the
SM prediction. 

\section{Status of $B\to\tau\nu$}

The \btn\ branching fraction is given in the Standard Model by
\begin{eqnarray*}
\mathcal{B}(B^+\to\tau^+\nu) = \frac{G^2_Fm_Bm_\tau^2}{8\pi}
   \left[1-\frac{m_\tau^2}{m_B^2}\right]^2 \tau_{B^+}f_B^2|V_{ub}|^2,
\end{eqnarray*}
where $G_F$ is the Fermi constant, $\tau_{B^+}$ is the $B^+$ lifetime
and $m_B$ and $m_\tau$ are the $B^+$ meson and $\tau$ lepton masses. 
The branching fraction can be significantly modified by New Physics
containing a charged Higgs boson.  In minimal supersymmetric
models, for example, the SM branching fraction is modified~\cite{btn-theory}:
\begin{eqnarray*}
\lefteqn{\mathcal{B}(B^+\to\tau^+\nu) = } \\
& & \mathcal{B}(B^+\to\tau^+\nu)_{SM}
  \left[1-\tan^2\beta\frac{m^2_{B^+}}{m^2_{H^+}}\right]^2.
\end{eqnarray*}

Belle and \babar\ have measured the branching fraction of
\btn~\cite{btn-babar1,btn-babar2,btn-belle}. The experimental
techniques are similar to the \bknn\ analyses: reconstruction of the
tag $B$-meson in either a semileptonic or hadronic decay mode,
multivariate discriminators to further suppress backgrounds and
construction of an $E_{\rm{extra}}$ variable, which represents the
energy left over after excluding daughters of the tag $B$-meson and the $\tau$
lepton, reconstructed in one of: $\mu^+\nu\overline{\nu}$,
$e^+\nu\overline{\nu}$, $\pi^+\nu\overline{\nu}$ or
$\pi^+\pi^0\nu\overline{\nu}$.  An excess near $E_{\rm{extra}} \sim 0$ is a
signature of \btn\ decays.

Results are shown in Table~\ref{tab:btn}. While no single measurement
is highly significant, the average is significant and consistent with
the Standard Model expectation.
\begin{table}[h]
\label{tab:btn}
\begin{center}
\caption{Results for \btn. The uncertainties on the branching fractions are statistical
and systematic.}
\begin{tabular}{lcc}\hline
Experiment & Tag  & Branching \\
           & Mode & Fraction \\ \hline
BaBar~\cite{btn-babar1} & had     & $(1.8\pm0.9\pm0.5)\times 10^{-4}$  \\
BaBar~\cite{btn-babar2}  & SL      & $(0.9\pm0.6\pm0.1)\times 10^{-4}$  \\
Belle~\cite{btn-belle} & had     & $(1.8\pm0.5\pm0.5)\times 10^{-4}$  \\ 
\hline
HFAG Average & - & $(1.41 \pm 0.42) \times 10^{-4}$ \\
\hline
\end{tabular}
\end{center}
\end{table}

\section{Conclusions}

Radiative penguin and leptonic $B$-meson decays are excellent probes
for investigating the effects of New Physics. Although current
measurements are in agreement with the Standard Model expectations,
they are still quite useful for setting bounds on possible NP
models. The \bsg\ and \btn\ measurements, for example, put strong
constraints on the mass of charged Higgs bosons in Type II two-Higgs
double models~\cite{haisch1}.  The \bsg\ branching fraction
measurements also constrain models with universal extra
dimensions~\cite{haisch2}.

Finally, I would like to express my heartfelt thanks to the conference organizers for 
a stimulating conference and warm hospitality.

\end{document}